\documentclass[3p,11pt,authoryear]{elsarticle}

\usepackage{xcolor,hyperref}
\hypersetup{
   colorlinks,
   linkcolor={blue!50!black},
   citecolor={blue!50!black},
   urlcolor={blue!80!black}
}

\usepackage{amsfonts}
\usepackage{amsmath}
\usepackage{amssymb}
\usepackage{amsthm}
\usepackage{mathrsfs}
\usepackage{graphics}
\usepackage{lscape}
\usepackage{changebar}
\usepackage{graphicx}
\usepackage{epsfig}
\usepackage{bm}
\usepackage{color}
\newcommand{\tr}{\hbox{tr}}

\newcommand{\B}[1]{{\bm{#1}}}

\renewcommand{\=}{\!=\!}

\journal{Journal of the Mechanics and Physics of Solids}

\begin{document}

\begin{frontmatter}

\title{Theory of unconventional singularities of frictional shear cracks}

\author{Efim A. Brener$^{1,2}$ and Eran Bouchbinder$^3$\footnote{Corresponding author. E-mail address: \url{eran.bouchbinder@weizmann.ac.il}}}
\address{$^1$Peter Gr\"unberg Institut, Forschungszentrum J\"ulich, D-52425 J\"ulich, Germany\\
$^2$Institute for Energy and Climate Research, Forschungszentrum J\"ulich, D-52425 J\"ulich, Germany\\
$^3$Chemical and Biological Physics Department, Weizmann Institute of Science, Rehovot 7610001, Israel}
\date{\today}
\begin{abstract}
Crack-like objects that propagate along frictional interfaces, i.e.~frictional shear cracks, play a major role in a broad range of frictional phenomena. Such frictional cracks are commonly assumed to feature the universal square root near-edge singularity of ideal shear cracks, as predicted by Linear Elastic Fracture Mechanics. Here we show that this is not the generic case due to the intrinsic dependence of the frictional strength on the slip rate, even if the bodies forming the frictional interface are identical and predominantly linear elastic. Instead, frictional shear cracks feature unconventional singularities characterized by a singularity order $\xi$ that differs from the conventional $-\tfrac{1}{2}$ one. It is shown that $\xi$ depends on the friction law, on the propagation speed and on the symmetry mode of loading. We discuss the general structure of a theory of unconventional singularities, along with their implications for the energy balance and dynamics of frictional cracks. Finally, we present explicit calculations of $\xi$ and the associated near-edge fields for linear viscous-friction --- which may emerge as a perturbative approximation of nonlinear friction laws or on its own --- for both in-plane (mode-II) and anti-plane (mode-III) shear loadings.
\end{abstract}

\begin{keyword}
Friction, Cracks, Singularity, Elastodynamics, Energy balance
\end{keyword}

\end{frontmatter}


\section{Background and motivation}
\label{sec:intro}

A distinguishing feature of cracks in a broad range of materials and physical situations is the emergence of singular fields (e.g.~stress, strain and particle velocity) near their edges. These singularities play important roles in determining the physical properties and dynamics of many natural and man-made systems, and consequently attract considerable interest. The most well-known and widely-used crack singularity emerges in the classical Linear Elastic Fracture Mechanics (LEFM) theory~\citep{Freund1998,Broberg1999}. In this framework, the linearized field theory of elasticity is assumed to hold and power-law solutions proportional to $r^\xi$, with $\xi\=-\tfrac{1}{2}$ and where $r$ is the distance from the crack edge, exist for traction-free boundary conditions along the crack surfaces~\citep{Freund1998,Broberg1999}. Under a broad range of physical conditions, sometimes termed small-scale-yielding conditions, the LEFM singular fields dominate the mechanical state of the material over some spatial range. As such, and despite that these singular fields are inevitably regularized at the smallest scales near the crack edge, they have profound implications for crack initiation and dynamics.

The $\xi\=-\tfrac{1}{2}$ singularity is associated with a finite influx of energy per unit area $G$ into the edge region, which is uniquely determined by the intensity of the singularity (the so-called stress intensity factor $K$), independently of the way the singularity is being regularized. The intensity of the singularity $K$ encapsulates information about the large scales of a given fracture problem, in particular about the loading configuration and the external geometry of the stressed body; once $K$ is known, the singular fields set the effective boundary conditions for the small-scale nonlinear/dissipative problem near the edge, where fracture is actually taking place. This scale separation picture is at the heart of conventional fracture theory. In what follows, we refer to the celebrated square root LEFM singularity, $\xi\=-\tfrac{1}{2}$, as the conventional crack singularity.

Singular crack edge fields that feature $\xi\!\ne\!-\tfrac{1}{2}$ emerge in certain classes of nonlinear materials, such as power-law hardening plastic materials (e.g.~the well-known HRR singularity~\citep{Hutchinson1968,Rice1968plane}) and nonlinear elastic materials~\citep{Bouchbinder.08,Bouchbinder.14,Long2015crack,Long2021review}. Such nonlinear field theories will not be discussed in this paper. Singular crack edge fields with $\xi\!\ne\!-\tfrac{1}{2}$ are known to emerge, even in the framework of the linearized field theory of elasticity and for traction-free boundary conditions along the crack surfaces, in the presence of material contrast across the fracture plane. Most notably, it has been shown~\citep{Williams1959stresses,RiceSih1965,Rice1988interfacial,Comninou1990overview} that cracks located along an interface between two dissimilar linear elastic bodies feature an oscillatory square root singularity of the form $\xi\=-\tfrac{1}{2}+i\varrho$. The imaginary part $\varrho$, which depends on the material contrast and vanishes in its absence, is responsible for the oscillatory nature of this singularity. In general, though, far less is known about near-edge singularities that differ from $\xi\=-\tfrac{1}{2}$ in homogeneous linear elastic materials (i.e.~in the absence of any material contrast). Such singularities are termed hereafter unconventional singularities.

The physical origin of such unconventional singularities is processes that take place at the crack surfaces, which consequently affect the boundary conditions for the linear elastic bulk along these surfaces. Generically, such physical processes are described by a boundary relation between various fields, rather than by prescribing the fields themselves. While such physical processes might be relevant to various classes of problems, mainly --- but not exclusively --- those involving frictional shear cracks, to the best of our knowledge only very few unconventional singularities are discussed in the literature. One such example concerns tensile cracks in hydraulic fracture~\citep{Desroches1994}. In this case, crack propagation in an impermeable linear elastic solid is hydraulically driven by a power-law fluid and asymptotic singular solutions with $\xi\=-\tfrac{n}{n+2}$ emerge, where $n$ is the exponent characterizing the power-law fracturing fluid. Other examples emerge in the context of frictional shear cracks, i.e.~shear cracks whose surfaces interact under contact, which is the focus of this paper.

When frictional shear cracks are considered, the contact interaction between the sliding crack surfaces is described by an interfacial relation between the frictional strength (stress) $\tau$, the slip rate $v$ and the net normal stress (the difference between the solid interfacial normal stress and the pressure of the pore fluid, if the latter exists). In the context of earthquake faults, where pore fluids exist inside fault gouges, it has been shown that an unconventional singularity with $\xi\=-\tfrac{1}{4}$ emerges in the framework of a nonlinear thermal pressurization model~\citep{Viesca2015}. Unconventional singularities emerge also in frictional shear cracks where $\tau$ is linearly related to $v$ in a viscous-like relation~\citep{Ida1974slow,Brener2002,Brener2005,Brener2020unconventional}. Such a linear viscous-friction relation, under quasi-static conditions, has been shown to lead to $\xi\!>\!-\tfrac{1}{2}$~\citep{Ida1974slow}, where the exact order of singularity depends on both the viscous-friction coefficient and on the crack propagation speed (in this case in the slow, quasi-static regime). Later, the leading order correction to $\xi\=-\tfrac{1}{2}$ has been derived for frictional shear cracks propagating along a linear viscous-friction interface separating a linear elastic solid and a rigid substrate (i.e.~in the presence of an infinite bi-material contrast), both in the quasi-static and fully elastodynamic regimes~\citep{Brener2002}. Finally, very recently unconventional singularities have been shown to exist near the edge of cracks propagating along nonlinear rate-and-state frictional interfaces, where linear viscous-friction emerges as a perturbative approximation~\citep{Brener2020unconventional}.

With this background in mind, the purpose of this paper is 3-fold; first, we aim at highlighting the rather generic existence of unconventional singularities that emerge when the boundary conditions along the crack surfaces are given as a relation between various fields. Second, we aim at discussing the quite significant physical implications of unconventional singularities for frictional dissipation and rupture energy balance. These two goals are addressed in Sect.~\ref{sec:main}. Third, we aim at presenting analytic results for the unconventional singularities of elastodynamic frictional shear cracks characterized by a linear viscous-friction relation. The latter is relevant, for example, for interfaces composed of gouge layers/granular materials under fast shearing~\citep{Kuwano2013crossover} or for hydrodynamically lubricated interfaces~\citep{Lu2006stribeck}, or as a perturbative approximation for other nonlinear friction laws~\citep{Brener2020unconventional}. This is achieved in Sect.~\ref{sec:viscous}, where $\xi\!>\!-\tfrac{1}{2}$ is shown to continuously depend on the viscous-friction coefficient, on the crack propagation speed and on the symmetry mode of loading, i.e.~in-plane (mode-II) or anti-plane (mode-III) shear. Section~\ref{sec:viscous} also presents the azimuthal dependence, i.e.~away from the frictional interface, of the asymptotic unconventional singular fields for mode-II and mode-III cracks. These results provide yet another concrete example for the emergence of unconventional crack singularities and their generic properties. Finally, some discussion and concluding remarks are offered in Sect.~\ref{sec:summary}.

\section{The emergence of unconventional singularities in frictional cracks and their implications for rupture energy balance}
\label{sec:main}

Consider a frictional shear crack propagating steadily at a velocity $c_{\rm r}\!>\!0$ along an interface. In a fixed Cartesian coordinate system $(x,y,z)$, the interface is defined by the $y\=0$ plane and the crack propagates in the positive $x$ direction. The bodies that form the contact interface are identical and satisfy some bulk field equations, accompanied by some boundary conditions along the crack surfaces, and subjected to some external loading. The external loading and the geometry of the bodies in contact are such that the emerging fields feature a well-defined symmetry in the $z$ direction, perpendicular to the $x\!-\!y$ plane. In particular, we assume that the resulting solutions are effectively two dimensional, i.e.~the displacement vector field is ${\B u}(x,y,t)$ and the stress tensor field is ${\B \sigma}(x,y,t)$. The interfacial shear stress, either $\sigma_{xy}(x,y\=0,t)$ for in-plane shear or $\sigma_{yz}(x,y\=0,t)$ for anti-plane shear, is continuous across the interface and equals the frictional strength $\tau(x)$. The interfacial displacement field is discontinuous across the interface, where the magnitude of discontinuity is $\delta(x,t)$, i.e.~the slip displacement. For in-plane (mode-II) shear loading, where ${\B u}(x,y,t)\=(u_x(x,y,t), u_y(x,y,t),0)$, one has $\delta(x,t)\!\equiv\!{u}_x(x,y\!\to\!0^+,t)-{u}_x(x,y\!\to\!0^-,t)$, where $+/-$ correspond to the upper/lower bodies, respectively. For anti-plane (mode-III) shear loading, where ${\B u}(x,y,t)\=(0,0,u_z(x,y,t))$, one has $\delta(x,t)\!\equiv\!{u}_z(x,y\!\to\!0^+,t)-{u}_z(x,y\!\to\!0^-,t)$. The most relevant physical field for frictional dynamics is the slip velocity $v(x,t)\=\dot\delta(x,t)$, where the superposed dot corresponds to a partial time derivative.

\subsection{General theoretical considerations}
\label{subsec:theory}

\begin{figure}[ht!]
\centering
\includegraphics[width=0.65\textwidth]{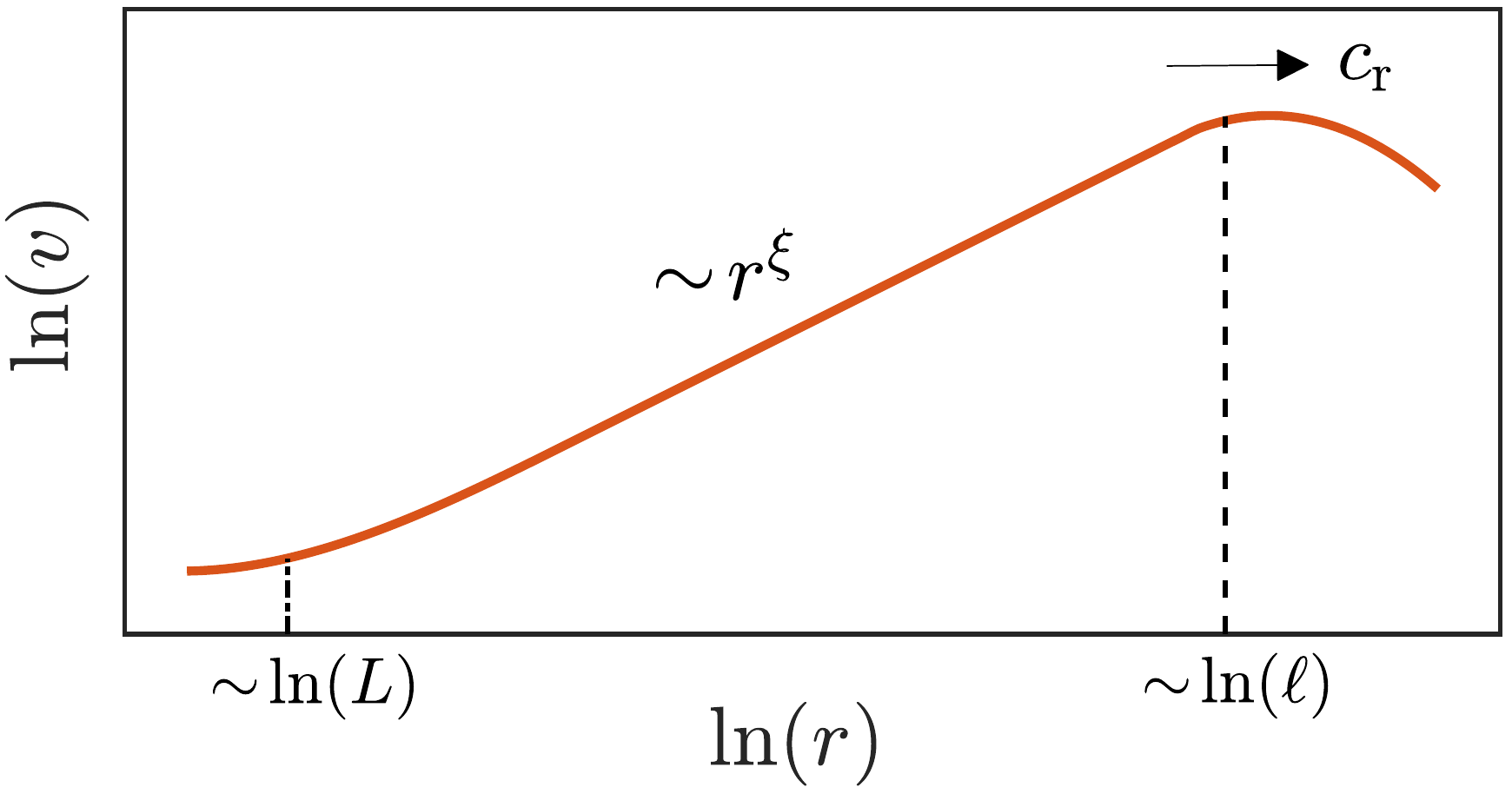}
\caption{A sketch of the slip velocity field $v(r)$ left behind a frictional crack propagating at an instantaneous velocity $c_{\rm r}$, where $(r,\theta)$ is co-moving polar coordinate system, with $r$ being the distance from the crack edge and $\theta\!=\!0$ is the propagation direction. If singular solutions, $v(r)\!\sim\!r^\xi$ with $\xi\!<\!0$, exist, they appear at an intermediate region $\ell\!\ll\!r\!\ll\!L$, where $\ell$ is a localization/regularization length near the crack edge (sometimes termed the cohesive zone size) and $L$ is a macroscopic scale (e.g.~the crack length), both are schematically marked on the figure (note that $r$ increases from right to left). See text for additional discussion.}
\label{fig:sketch}
\end{figure}

Our main interest is in the possible emergence of singular solutions for $v(x,t)$ and $\tau(x,t)$ in the vicinity of the steadily propagating crack edge. It would be useful to consider a co-moving polar coordinate system $(r,\theta)$, where $r$ is the distance from the edge and $\theta\=0$ is the propagation direction, cf.~Fig.~\ref{fig:sketch}. Formally, the singular solutions of interest may emerge as intermediate asymptotics, i.e.~for $\ell\!\ll\!r\!\ll\!L$, where $\ell$ is a localization/regularization length near the crack edge (sometimes termed the cohesive zone size) and $L$ is a macroscopic scale (e.g.~the crack length), cf.~Fig.~\ref{fig:sketch}. The lengths $\ell$ and $L$, with the condition $\ell\!\ll\!L$, allow to discuss in more quantitative terms the scale separation picture commonly associated with cracks. Intermediate asymptotic solutions can, in principle, be approached in two ways; one can formulate the so-called ``outer problem'', where the characteristic scale is $L$ and the singular behavior emerges in the limit $r/L\!\ll\!1$. Alternatively, one can formulate the so-called ``inner problem'' (sometimes also termed the ``boundary layer approximation''), where the characteristic scale is $\ell$ and the singular behavior emerges in the $r/\ell\!\gg\!1$ limit. Most of the discussion below will adopt the former approach, yet one example will follow the latter. It is important to note that singular power-law solutions do not always exist and that even if they do, the singularity order may be different for the slip velocity $v(r)$ and for the stress $\tau(r)$. Yet, whenever the bodies forming the contact interface are predominantly linear elastic, some constraints emerge, as we discuss next.

As stated in Sect.~\ref{sec:intro}, we focus in this paper on the linearized field theory of elasticity, i.e.~we assume that the bodies forming the contact interface are predominantly linear elastic. In this case, the bulk linear elastodynamic field equations impose an interfacial steady-state relation between $\tau(r)$ and $v(r)$ of the form~\citep{Weertman1980}
\begin{equation}
\label{eq:boundary_integral}
\tau(r) \sim \frac{\mu}{c_s} \int_0^\infty \!\frac{v(s)}{s-r}\,ds \ ,
\end{equation}
where $\mu$ is the shear modulus, $c_s$ is the shear wave speed and the unspecified pre-factor is a dimensionless function of $c_{\rm r}/c_s$, which depends on the symmetry mode of loading (see Sect.~\ref{sec:viscous}). Note that while Eq.~\eqref{eq:boundary_integral} involves the linear elastodynamic Green's function of infinite domains, it remains valid for our purposes here independently of the geometry of the macroscopic bodies in contact, as we focus on intermediate asymptotic solutions.

Equation~\eqref{eq:boundary_integral} implies that if $\tau(r)$ and $v(r)$ feature a singular behavior, their singularities are related. In particular, using the identity $\int_0^\infty s^\xi(s-r)^{-1}ds\=-\pi \cot(\xi\pi)r^\xi$, it implies~\citep{King2009hilbert}
\begin{equation}
\label{eq:linear}
v(r) \sim r^\xi \qquad\hbox{and}\qquad \tau(r) \sim \cot(\xi\pi)\,r^\xi
\end{equation}
for $\theta\=\pm\pi$, i.e.~along the interfacial region left behind the propagating frictional crack, cf.~Fig.~\ref{fig:sketch}. Equation~\eqref{eq:linear} shows that in the framework of bulk linear elasticity, if a singularity ($\xi\!<\!0$) emerges, either both $\tau(r)$ and $v(r)$ feature the same singularity order $\xi$ or $\tau(r)$ is non-singular (the latter is the case if $\cot(\xi\pi)\!=\!0$). Moreover, and most importantly for our general argument, Eq.~\eqref{eq:linear} shows that whether a singularity emerges at all and if so, what its order $\xi$ is, is determined by the interfacial constitutive relation. In more mathematical terms, it is determined by the boundary conditions imposed on the bulk field equations along the contact interface.

Generally, an interfacial constitutive relation takes the form $\tau(v,...)$, where the dots stand for additional fields such as pore fluid pressure, temperature and internal state fields that characterize the structural state of the interface, all which can satisfy their own evolution equations~\citep{Ruina1983,Rice1983,Marone1998a,Baumberger2006}. As is evident from Eq.~\eqref{eq:linear}, and as will be further discussed below, the most important --- and minimal --- ingredient for the emergence of unconventional singularities is the slip rate $v$ dependence of the frictional strength $\tau$. In the LEFM case of ideal cracks, such rate dependence does not exist and $\tau$ is assumed to vanish, leading to traction-free boundary conditions. Consequently, the boundary condition in this case is satisfied by selecting $\xi\!=\!-\tfrac{1}{2}$ in Eq.~\eqref{eq:linear}, ensuring $\tau(r)\=0$. The same applies to the boundary condition $\tau(r)\=\tau_{\rm res}$, where a constant residual strength $\tau_{\rm res}$ is reached over a small localization/regularization length $\ell$ near the crack edge, as assumed in the widely-used slip-weakening friction model~\citep{Palmer1973,Ida1972}. The reason that the finite constant $\tau_{\rm res}$ for $r\!>\!\ell$ does not affect the singularity order is that we consider here the ``outer problem'', where $r$ in Eq.~\eqref{eq:boundary_integral} in fact stands for $r/L$ and the $r/L\!\to\!0$ limit is considered. That is, the constant $\tau_{\rm res}$ is not accounted for by the leading order solution in Eq.~\eqref{eq:linear}, but rather by higher order, sub-leading contributions. $\xi\!=\!-\tfrac{1}{2}$ is nothing but the conventional singularity case, where $\tau(r)$ behind the crack edge is constant and the slip velocity $v(r)$ features there the $\xi\!=\!-\tfrac{1}{2}$ singularity.

Unconventional singularities may emerge, as stated above, when $\tau$ depends on $v$, which is the generic case in frictional constitutive relations~\citep{Ruina1983,Rice1983,Marone1998a,Baumberger2006}. The simplest scenario would be when $\tau$ is linear in $v$, i.e.~$\tau(v)\=\tau_{\rm res}+\eta\,v$, where $\eta$ is a linear viscous-friction coefficient. This interfacial constitutive relation, which will be studied in Sect.~\ref{sec:viscous}, may emerge in various physical situations, either as a complete constitutive law in a certain range of slip velocities~\citep{Lu2006stribeck,Kuwano2013crossover} or as a perturbative approximation for a nonlinear constitutive law~\citep{Brener2020unconventional}. In this case, Eq.~\eqref{eq:linear} implies that the unconventional singularity order $\xi\!\ne\!-\tfrac{1}{2}$ depends on the viscous-friction coefficient $\eta$ and on the yet-unspecified pre-factor function of $c_{\rm r}/c_s$ in Eq.~\eqref{eq:boundary_integral}, as will be discussed in Sect.~\ref{sec:viscous}. Here, again, the constant $\tau_{\rm res}$ does not contribute to the leading order solution in the $r/L\!\to\!0$ limit. Finally, note that Eq.~\eqref{eq:linear} also implies that for unconventional singularities that slightly deviate from the conventional $\xi\=-\tfrac{1}{2}$ one, the amplitude of the $\tau(r)$ singularity is much smaller than that of $v(r)$.

It is clear from the above discussion that not all interfacial constitutive relations would give rise to the emergence of singular power law fields, assuming that Eq.~\eqref{eq:linear} holds. For example, consider the nonlinear friction law $\tau(v)\!\sim\!v^2$. This relation is obviously inconsistent with Eq.~\eqref{eq:linear}, which simply implies that singular solutions do not emerge for this nonlinear friction law (again, assuming that Eq.~\eqref{eq:linear} holds). Other nonlinear friction laws, however, might give rise to unconventional singularities. Recently, such an example was given in~\citet{Viesca2015}, where interfaces separating linear elastic bodies have been considered and the ``inner problem'' has been studied. That is, Eqs.~\eqref{eq:boundary_integral}-\eqref{eq:linear} remain valid, but this time $r$ stands for $r/\ell$ and singular power-law solutions are sought for in the $r/\ell\!\to\!\infty$ limit (as opposed to the $r/L\!\to\!0$ limit of the same equations discussed above). The constitutive law considered in~\citet{Viesca2015} that is relevant for the ``inner problem'' is the so-called slip-on-a-plane law~\citep{Rice2006}. The latter is a nonlinear friction law of the form $\tau(r)-\tau_0\!\sim\!-\!\int_0^r\!\tau(s)v(s)(r-s)^{-1/2}ds$, with a constant $\tau_0$, which emerges in the drained, large-slip limit of a thermal pressurization model~\citep{Rice2006}. Upon substituting Eq.~\eqref{eq:linear} inside the slip-on-a-plane constitutive equation and looking at the $r\!\to\!\infty$ limit, one obtains $\xi\!=\!-\tfrac{1}{4}$. In this case, the nonlinear integral term $\int_0^r\!\tau(s)v(s)(r-s)^{-1/2}ds$ asymptotically balances the constant $\tau_0$, which is a leading order contribution in the $r\!\to\!\infty$ limit.

To sum up, we have shown that when the bodies forming the frictional interface are predominantly linear elastic such that Eq.~\eqref{eq:boundary_integral} is valid, frictional shear cracks generically feature unconventional singular solutions, i.e.~Eq.~\eqref{eq:linear} with $\xi\!\ne\!-\tfrac{1}{2}$, when the frictional strength $\tau$ depends on the slip velocity $v$. In this case, the singularity order $\xi$ is non-universal, rather it depends on the friction law, on the crack propagation speed and on the symmetry mode of loading, as will be further demonstrated below. We note that solutions to the interfacial integral problem defined in Eq.~\eqref{eq:boundary_integral} allow the calculation of the singularity order $\xi$, but not of the full field distribution near the propagating crack edge. These unconventional asymptotic fields will be derived in Sect.~\ref{sec:viscous}. Next, we discuss some general physical implications of the existence of unconventional singularities, most notably for rupture energy balance.

\subsection{Implications for rupture energy balance}
\label{subsec:budget}

One of the hallmarks of the conventional singularity $\xi\=-\tfrac{1}{2}$ is that it is accompanied by a finite influx of elastic energy $G$ (energy per unit area, the so-called energy release rate) into the edge region. $G$ is commonly evaluated through a closed contour integral, termed the J-integral~\citep{Rice1968,Freund1998}, which is path independent as long as the selected path resides inside the singular region~\citep{Freund1998}. $G$ is dissipated near the edge of the crack over the localization length $\ell$, where dissipation is quantified by the fracture energy $G_{\rm c}$. Consequently, the edge-localized energy balance $G\=G_{\rm c}$ serves as an equation of motion for the crack (if its trajectory is a priori known, e.g.~if it propagates along a predetermined interface), whose solution allows to determine the crack propagation speed $c_{\rm r}$. Our main goal here is to understand the ways in which this rupture energy balance changes in the presence of unconventional singularities

To set the stage for the discussion of the implications of unconventional singularities, let us consider the following dissipation integral~\citep{Bizzarri2010a,Brener2020unconventional}
\begin{equation}
\label{eq:E_BD}
E_{_{\rm BD}}\!(r)=c_{\rm r}^{-1}\!\!\int_0^r\!\left[\tau(s)-\tau_{\rm res} \right] v(s)\,ds \ .
\end{equation}
$E_{_{\rm BD}}\!(r)$, commonly termed the breakdown energy (hence the subscript), quantifies the dissipation on top of the background frictional dissipation (heat) associated with sliding against the residual stress $\tau_{\rm res}$ (in case the latter is finite and well defined~\citep{Barras2019emergence}). That is, $E_{_{\rm BD}}\!(r)$ is the dissipation associated with rupture propagation. Note that the integration in Eq.~\eqref{eq:E_BD} extends from the crack edge to the region left behind it and that $c_{\rm r}$ is assumed to be time independent. In the case of conventional cracks, Eq.~\eqref{eq:linear} holds with $\xi\=-\tfrac{1}{2}$ for $r\!>\!\ell$, and more importantly for our present discussion, the stress $\tau(r)$ drops to the residual stress $\tau_{\rm res}$ (be it finite or zero) over the localization/regularization length $\ell$. Consequently, $E_{_{\rm BD}}\!(r)\=G_{\rm c}$ for $r\!>\!\ell$ (i.e.~independently of $r$), where $G_{\rm c}\!\equiv\!c_{\rm r}^{-1}\!\int_0^\ell[\tau(s)-\tau_{\rm res}] v(s) ds$.

What form does $E_{_{\rm BD}}\!(r)$ take in the presence of unconventional singularities? Using the just-stated definition of the fracture energy $G_{\rm c}$, and using Eq.~\eqref{eq:linear} with $\xi$ in the range $-1\!<\!\xi\!<\!0$, one obtains
\begin{equation}
\label{eq:E_BD1}
E_{_{\rm BD}}\!(r)=G_{\rm c} \left(r/\ell\right)^{1+2\xi} \ ,
\end{equation}
for $r\!\ge\!\ell$. For the conventional singularity, $\xi\=-\tfrac{1}{2}$, the $r$-independent $E_{_{\rm BD}}\=G_{\rm c}$ result discussed above is recovered, justifying the LEFM-like equation of motion $G\=G_{\rm c}$. For any $\xi\!\ne\!-\tfrac{1}{2}$, $E_{_{\rm BD}}\!(r)$ features $r$ dependence, which in turn implies that the conventional scale separation picture --- where all of the dissipation associated with crack propagation is given by the fracture energy $G_{\rm c}$ on a length $\ell$ --- is no longer strictly valid. For $-\tfrac{1}{2}\!<\!\xi\!<\!0$, one has $1+2\xi\!>\!0$ and $E_{_{\rm BD}}\!(r)$ in Eq.~\eqref{eq:E_BD1} is an increasing function of $r\!>\!\ell$. That is, $E_{_{\rm BD}}\!(r)$ surpasses $G_{\rm c}$ due to a spatially-extended contribution to the dissipation. The saturation of $E_{_{\rm BD}}\!(r)$ is related to the length scale over which $\tau(r)$ decays to the residual stress $\tau_{\rm res}$. This length scale is typically macroscopic in nature, i.e.~much larger than $\ell$, as recently discussed in~\citet{Barras2020,Brener2020unconventional}. The fact that the saturation of $E_{_{\rm BD}}\!(r)$ occurs over a macroscopic length much larger than $\ell$ implies that even for small deviations of $\xi$ from $-\tfrac{1}{2}$, $E_{_{\rm BD}}\!(r)$ can deviate quite substantially from $G_{\rm c}$~\citep{Barras2020,Brener2020unconventional}.

For $-1\!<\!\xi\!<\!-\tfrac{1}{2}$, one has $1+2\xi\!<\!0$ and $E_{_{\rm BD}}\!(r)$ in Eq.~\eqref{eq:E_BD1} is a decreasing function of $r\!>\!\ell$. That is, $E_{_{\rm BD}}\!(r)$ drops below $G_{\rm c}$, again due to the intervention of macroscopic length scales much larger than $\ell$, which breaks the conventional scale separation in cracks. These results raise various basic questions regarding frictional rupture energy balance: Under what conditions $G_{\rm c}$ can be meaningfully separated from $E_{_{\rm BD}}$? Under what conditions the influx of elastic energy $G$ is well-defined? Most importantly, under what conditions an approximate edge-localized equation of motion of the form $G\!\approx\!G_{\rm c}$, which allows to determine the crack propagation speed, remains valid? In addressing these questions, one can roughly refer to two regimes. First, in physical situations in which $\xi$ deviates from $-\frac{1}{2}$ slightly, all of the aforementioned quantities remain well-defined and $G\!\approx\!G_{\rm c}$ remains approximately valid; in fact, a perturbative approach that employs $\xi\=-\frac{1}{2}$ as the leading order solution can be systematically developed, as demonstrated recently in~\citet{Brener2020unconventional}. It is important to stress again that small deviations from $\xi\=-\frac{1}{2}$ do not necessarily imply small deviations of $E_{_{\rm BD}}$ from $G_{\rm c}$~\citep{Barras2020,Brener2020unconventional}.

In physical situations in which $\xi$ deviates from $-\frac{1}{2}$ significantly, qualitative changes in the crack propagation energy balance emerge. To understand these qualitative changes, we need to discuss in some more detail the implications of unconventional singularities also for the elastic energy influx $G$. Scaling-wise, $G$ is determined by a contour integral over the linear elastic energy density, i.e.~$G\!\sim\!\int {\B \sigma}\!:\!\nabla{\B u}\,r\,d\theta$~\citep{Freund1998}. Equation~\eqref{eq:linear} implies that the stress ${\B \sigma}$ and the displacement gradient follow unconventional singular solutions ${\B \sigma}\!\sim\!K^{(\xi)} r^\xi$ and $\nabla{\B u}\!\sim\!K^{(\xi)} r^\xi/\mu$ (where the tensorial nature of these fields is not considered here, the focus is on the scaling structure). Here $K^{(\xi)}$ is the $\xi$-generalized stress intensity factor, whose dimension is stress $\times$ length$^{-\xi}$. Plugging these into the scaling expression for $G$, one obtains
\begin{equation}
\label{eq:G_generalized}
G(r) \sim \frac{\left[K^{(\xi)}\right]^2 r^{1+2\xi}}{\mu} \ ,
\end{equation}
for $r\!\ge\!\ell$. This result is qualitatively different from the conventional singularity case. In the latter, where $\xi\=-\tfrac{1}{2}$, $G(r)$ is scale independent, i.e.~independent of the distance $r$ from the crack edge where the contour integration is performed. For $\xi\!\ne\!-\tfrac{1}{2}$, $G(r)$ is scale dependent, i.e.~it explicitly depends on $r$. This qualitatively new physics has a marked quantitative effect when $\xi$ significantly deviates from $-\tfrac{1}{2}$.

Crack propagation energy balance on a scale $r\!\ge\!\ell$ can still be formulated using Eqs.~\eqref{eq:E_BD1}-\eqref{eq:G_generalized}, taking the form
\begin{equation}
\label{eq:EOM_generalized}
G(r) = E_{_{\rm BD}}\!(r) \ ,
\end{equation}
for $r\!\ge\!\ell$. Plugging Eqs.~\eqref{eq:E_BD1}-\eqref{eq:G_generalized} into Eq.~\eqref{eq:EOM_generalized}, we observe that while both $G(r)$ and $E_{_{\rm BD}}\!(r)$ depend on $r$, this dependence is identical (i.e.~$\sim r^{1+2\xi}$), implying that Eq.~\eqref{eq:EOM_generalized} is in fact approximately $r$ independent, taking the form
\begin{equation}
\label{eq:EOM_generalized1}
G_{\rm c} \sim \frac{\ell^{^{1+2\xi}}\left[K^{(\xi)}\right]^2}{\mu} \ .
\end{equation}
Note that while formally this relation holds for $r\!\gg\!\ell$, where the singular solutions in Eq.~\eqref{eq:linear} are ``well-developed'', for our purposes here we consider it valid for $r\!\ge\!\ell$, as done in Eq.~\eqref{eq:E_BD1} (in fact, this relation has been implicity assumed to hold in deriving Eq.~\eqref{eq:E_BD1}). Equation~\eqref{eq:EOM_generalized1} marks a sharp deviation from conventional fracture theory; it shows that crack propagation energy balance explicitly involves the localization/regularization length $\ell$. That is, unlike the conventional $\xi\=-\tfrac{1}{2}$ case, where energy balance $G_{\rm c}\!\sim\!K^2/\mu$ (here $K\!\equiv\!K^{(-{\scriptstyle 1/2})}$) involves the fracture energy $G_{\rm c}$ without specifying the length over which it is defined, Eq.~\eqref{eq:EOM_generalized1} shows that for $\xi\!\ne\!-\tfrac{1}{2}$, the length $\ell$ over which the dissipation associated with $G_{\rm c}$ takes place should be explicitly specified. Most notably, it implies that when $\xi$ significantly deviates from $-\tfrac{1}{2}$, the localization/regularization length $\ell$ plays a role, not just the dissipation $G_{\rm c}$ over this length.

The discussion above, and in particular Eqs.~\eqref{eq:E_BD1} and~\eqref{eq:EOM_generalized1}, assumed that the fracture energy $G_{\rm c}$ can be meaningfully separated from $E_{_{\rm BD}}$ and that localization/regularization length $\ell$ can be identified. This would be the case if the dissipative physics for $r\!<\!\ell$ is quite different from those for $r\!>\!\ell$, which would manifest itself in a markedly different $r$ dependence of $E_{_{\rm BD}}\!(r)$ for $r\!<\!\ell$, compared to that of Eq.~\eqref{eq:E_BD1}. This has been recently demonstrated in~\citet{Brener2020unconventional}. A complementary way to probe the same physics, relevant for earthquake mechanics, would be to consider the following quantity~\citep{Palmer1973,Abercrombie2005,Tinti2005,Viesca2015,Nielsen2016}
\begin{equation}
\label{eq:Gf}
G_{\rm f}(\delta)\equiv \int_0^{\delta}[\tau(\delta')-\tau(\delta)]\,d\delta' \ .
\end{equation}
$G_{\rm f}(\delta)$ is clearly related to $E_{_{\rm BD}}\!(r)$ through the steady-state relation $v\=c_{\rm r}\,d\delta/dr$, with one difference; in the former $\tau(\delta)$ is used as the reference stress, while in the latter it is the constant residual stress $\tau_{\rm res}$~\citep{Barras2019emergence}. $G_{\rm f}(\delta)$ of Eq.~\eqref{eq:Gf} is extensively invoked in the geophysical literature, where seismological source spectra are used to estimate it~\citep{Abercrombie2005,Tinti2005,Viesca2015,Nielsen2016}. In contrast, while $E_{_{\rm BD}}\!(r)$ in Eq.~\eqref{eq:E_BD} is a physically transparent quantity that can be analytically calculated under some conditions (e.g.~as done in Eq.~\eqref{eq:E_BD1}) and can be evaluated using numerical simulations, it is not directly accessible using seismological observations.

Using $v\=c_{\rm r}\,d\delta/dr$ and Eq.~\eqref{eq:linear} to obtain $\tau(\delta)\!\sim\!\delta^{\scriptstyle\frac{\xi}{1+\xi}}$, and then evaluating the integral in Eq.~\eqref{eq:Gf}, results in
\begin{equation}
\label{eq:Gf1}
G_{\rm f}(\delta) = G_{\rm c}\left(\!\frac{\delta}{\delta_\ell}\!\right)^{\!\frac{1+2\xi}{1+\xi}} \ ,
\end{equation}
for $\delta\!\ge\!\delta_\ell$, where $\delta_\ell$ is defined such that $G_{\rm f}(\delta_\ell)\=G_{\rm c}$, i.e.~it is the slip displacement that corresponds to the length $\ell$. Note that similarly to Eq.~\eqref{eq:E_BD1}, for the conventional singularity $\xi\=-\tfrac{1}{2}$ we have $G_{\rm f}(\delta)\=G_{\rm c}$ for $\delta\!\ge\!\delta_\ell$. If $G_{\rm c}$ is indeed a distinct edge-localized quantity, then we expect $G_{\rm f}(\delta)$ for $\delta\!<\!\delta_\ell$ to be markedly different from Eq.~\eqref{eq:Gf1}, as has been recently demonstrated~\citep{Brener2020unconventional}. Otherwise, $G_{\rm c}$ cannot be meaningfully separated from $E_{_{\rm BD}}$ and crack dynamics are not controlled by local energy balance; rather, global energy considerations has to be invoked. The prediction that a well-defined fracture energy $G_{\rm c}$ implies that $G_{\rm f}(\delta)$ changes its behavior at some characteristic slip displacement $\delta_\ell$ may be useful in identifying $G_{\rm c}$ in seismological data~\citep{Abercrombie2005,Tinti2005,Viesca2015,Nielsen2016}, a potentially important implication.

In this section we developed a general theory of unconventional singularities in the framework of the linearized field theory of elasticity and discussed their quite far-reaching physical implications for rupture energy balance. We showed that unconventional singularities can introduce new physics compared to conventional fracture mechanics, entailing the introduction of concepts such as the breakdown energy which is different from the fracture energy, spatially-extended dissipation, scale dependent elastic energy influx and the explicit dependence on intrinsic dissipation length scales. In the next section we present asymptotic crack edge fields solutions in the presence of unconventional singularities, i.e.~going beyond the unconventional singularity order $\xi$ alone, and also present explicit calculations of $\xi$ for frictional shear cracks that are described by linear viscous-friction interfacial constitutive law.

\section{Unconventional singular fields and analytic results for linear viscous-friction}
\label{sec:viscous}

The linear elastic bulk problem has been accounted for in the previous section through the interfacial integral relation in Eq.~\eqref{eq:boundary_integral}. Here we first briefly discuss a complementary approach that starts directly from the bulk field equations, which allows the derivation of the relevant fields also away from the frictional interface. Then, we focus on deriving the unconventional singularity order $\xi$ for a linear viscous-friction constitutive relation. Linear elastic solids are described by the Navier-Lam\'e equation~\citep{Landau1986}
\begin{equation}
\nabla\cdot\B \sigma=(\lambda+\mu)\nabla\left(\nabla\cdot{\B u}\right)+\mu\nabla^2{\B u}~=~\rho\,\ddot{\B u} \ .
\label{Navier}
\end{equation}
Here $\B\sigma(x,y,t)$ and ${\B u}(x,y,t)$, which were already introduced above, satisfy Hooke's law $\B\sigma\=\lambda\,\tr \B\varepsilon\,\B I+ 2\mu\,\B\varepsilon$, where $\lambda$ is Lam\'e's first constant (the shear modulus $\mu$ was introduced earlier), $\B\varepsilon\=\tfrac{1}{2}[(\nabla \B u)+(\nabla \B u)^{\rm T}]$ is the linear strain tensor, $\B I$ is the identity tensor and $\rho$ is the mass density. We consider, as above, a shear crack propagating steadily at a velocity $c_{\rm r}\!>\!0$ in the positive $x$ direction.

Focus then on in-plane shear loading conditions (mode-II symmetry), for which the following symmetries $u_x(x,-y,t)\=-u_x(x,y,t)$ and $u_y(x,-y,t)\=u_y(x,y,t)$ are satisfied. These conditions imply that the crack surfaces are in contact, hence one should demand the continuity of the normal displacement, normal stress and shear stress behind the crack edge. We introduce a co-moving polar coordinate system $(r,\theta)$ according to $r_{d,s}e^{i\theta_{d,s}}\!\equiv\!x-c_{\rm r}t+i\alpha_{d,s}y$, where $c_d\=\sqrt{(\lambda+2\mu)/\rho}$ and $c_s\=\sqrt{\mu/\rho}$ are the dilatational and shear wave speeds, respectively, and $\alpha^2_{d,s}(c_{\rm r}) \!=\! 1\!-\!c_{\rm r}^2/c_{d,s}^2$. Consequently, we have $\tan(\theta_{d,s})\!=\!\alpha_{d,s}(c_{\rm r})\tan(\theta)$ and $r_{d,s}(c_{\rm r})\!=\!r\sqrt{1\!-\!(c_{\rm r}\sin(\theta)/c_{d,s})^2}$~\citep{Freund1998}. Using a standard complex-variable displacement potentials technique~\citep{Freund1998} and imposing the stated continuity conditions at $\theta\!=\!\pm\pi$, one obtains the following explicit solution
\begin{eqnarray}
\label{eq:mode_II}
u_x(r, \theta) &=& \frac{2K^{(\xi)}_{\rm II}}{\mu \sqrt{2\pi}D(c_{\rm r})} \left[2\alpha_s\,r_d^{\xi+1}\sin{\left[(\xi+1)\theta_d\right]}-\alpha_s\,(1+\alpha_s^2)\, r_s^{\xi+1}\sin{\left[(\xi+1)\theta_s\right]}\right],\nonumber\\
u_y(r,\theta) &=& \frac{2K^{(\xi)}_{\rm II}}{\mu \sqrt{2\pi}D(c_{\rm r})}
\left[2\alpha_s\,\alpha_d\,
r_d^{\xi+1}\cos{\left[(\xi+1)\theta_d\right]}-(1+\alpha_s^2)\,r_s^{\xi+1}\cos{\left[(\xi+1)\theta_s\right]}
\right] \ .
\end{eqnarray}
Here $D(c_{\rm r})\=4\alpha_s\alpha_d-(1+\alpha_s^2)^2$ is the Rayleigh function, $K^{(\xi)}_{II}$ quantifies the intensity of the singularity (already introduced in general terms in Sect.~\ref{subsec:budget}, cf.~Eq.~\eqref{eq:G_generalized}), i.e.~it is the analog of the mode-II stress intensity factor $K_{\rm II}$ that cannot be determined by the asymptotic analysis, and the singularity order $-1\!<\!\xi\!<\!0$ is yet to be determined. Note that nonsingular contributions in a series expansion in space can be obtained by increasing $\xi$ by integer multiples, and that the classical LEFM $K$-fields are recovered for $\xi\=-\tfrac{1}{2}$~\citep{Freund1998}.

Equations~\eqref{eq:mode_II} account for bulk linear elastodynamics, symmetries and continuity conditions across the crack surfaces, and are similar in their physical content to Eq.~\eqref{eq:boundary_integral} (the formal mathematical relation between the two will be briefly discussed below). As stressed in Sect.~\ref{sec:main}, the singularity order $\xi$ is determined by the constitutive relation behind the crack edge, i.e.~by the boundary conditions imposed on the crack surfaces. Here we consider a linear viscous-friction relation of the form
\begin{equation}
\label{eq:viscous_friction}
\tau(v)=\tau_{\rm res}+\eta\,v \ ,
\end{equation}
where $\eta$ is the linear viscous-friction coefficient. This constitutive relation is relevant, for example, for interfaces composed of gouge layers/granular materials under fast shearing~\citep{Kuwano2013crossover}, for hydrodynamically lubricated interfaces~\citep{Lu2006stribeck}, or as a perturbative approximation for other nonlinear friction laws~\citep{Brener2020unconventional}. Beyond its relevance for various physical systems and situations, Eq.~\eqref{eq:viscous_friction} is most useful for explicitly demonstrating our general argument as its linear nature allows analytic progress in calculating the singularity order $\xi$.

Equation~\eqref{eq:viscous_friction} is a relation between two fields behind the crack edge, the interfacial shear stress $\sigma_{xy}(r,\theta\!=\!\pi)$ that equals the frictional strength $\tau$ and the slip velocity $v(r)\=2\dot{u}_x(r,\theta\=\pi)$, both directly derivable from Eqs.~\eqref{eq:mode_II}. In particular, $\sigma_{xy}(r,\theta\!=\!\pi)$ is obtained by substituting Eqs.~\eqref{eq:mode_II} inside Hooke's law, i.e.~$\B\sigma\=\lambda\,\tr \B\varepsilon\,\B I+ 2\mu\,\B\varepsilon$ with $\B\varepsilon\=\tfrac{1}{2}[(\nabla \B u)+(\nabla \B u)^{\rm T}]$, and $v(r)$ is obtained using $\dot{u}_x(r,\theta\=\pi)\=-c_{\rm r}\partial_x u_x(r,\theta\=\pi)\=-c_{\rm r}\,\varepsilon_{xx}(r,\theta\=\pi)$. Substituting the results into Eq.~\eqref{eq:viscous_friction}, together with $\tau\=\sigma_{xy}(r,\theta\!=\!\pi)$, one obtains
\begin{equation}
\cot(\xi\pi) = - \frac{2\,{\vartheta}\,(c_{\rm r}/c_s)^3 \alpha_s(c_{\rm r})}{D(c_{\rm r})} \ ,
\label{eq:spectrum}
\end{equation}
where the dimensionless parameter
\begin{equation}
\label{eq:dimless_eta}
\vartheta\!\equiv\!\frac{\eta\,c_s}{\mu}
\end{equation}
quantifies the ratio between a velocity scale, $\mu/\eta$, emerging from the friction law ($\eta$) and bulk elasticity ($\mu$), and an elastic wave speed, $c_s$.

While Eq.~\eqref{eq:spectrum} is a transcendental equation that does not admit a closed-form solution, various analytic properties of it can be derived. First, since the right-hand-side of Eq.~\eqref{eq:spectrum} is negative, solutions for the singularity order $\xi(c_{\rm r}/c_s,\vartheta)$ exist in the range $-\tfrac{1}{2}\!\le\!\xi\!\le\!0$. Obviously, in the limit of vanishing viscous-friction, $\vartheta\!\propto\!\eta\!\to\!0$, the conventional singularity order $\xi\=-\tfrac{1}{2}$ is recovered, while for any finite $\eta$ an unconventional singularity emerges. Since the Rayleigh function $D(c_{\rm r})$ vanishes in the limit $c_{\rm r}\!\to\!c_{_{\rm R}}$, where $c_{_{\rm R}}$ is the Rayleigh wave speed, the right-hand-side of Eq.~\eqref{eq:spectrum} diverges and $\xi(c_{\rm r}/c_s,\vartheta)$ vanishes in this limit, i.e.~$\xi\!\to\!0$ (a similar behavior is expected for large $\vartheta$). In the opposite, quasi-static limit of $c_{\rm r}\!\ll\!c_{_{\rm R}}$, one obtains $D(c_{\rm r})\!\sim\!(c_{\rm r}/c_s)^2 + {\cal O}[(c_{\rm r}/c_s)^4]$, implying $\cot(\xi\pi)\!\propto\!-\eta\,c_{\rm r}/\mu$~\citep{Ida1974slow}. This result demonstrates that even in the quasi-static limit, $\xi(c_{\rm r}/c_s,\vartheta)$ may significantly differ from $-\tfrac{1}{2}$ due to the existence of an inertia-independent velocity scale $\mu/\eta$. Finally, for small deviations of $\xi$ from $-\tfrac{1}{2}$, one can define $\epsilon\!\equiv\!\xi+1/2\!\ll\!1$ and noting that $\cot(\xi\pi)\!\simeq\!-\pi(\xi + 1/2)\=-\pi \epsilon + {\cal O}(\epsilon^2)$, Eq.~\eqref{eq:spectrum} admits a leading order solution in the form $\epsilon\!\simeq\!2\,{\vartheta}\,(c_{\rm r}/c_s)^3 \alpha_s(c_{\rm r})/\pi D(c_{\rm r})$. This solution shows that the deviation of $\xi(c_{\rm r}/c_s,\vartheta)$ from $-\tfrac{1}{2}$ is linear in both $\vartheta\!\propto\!\eta$ and $c_{\rm r}$ (for small $c_{\rm r}/c_s$).

\begin{figure}[ht!]
\centering
\includegraphics[width=0.85\textwidth]{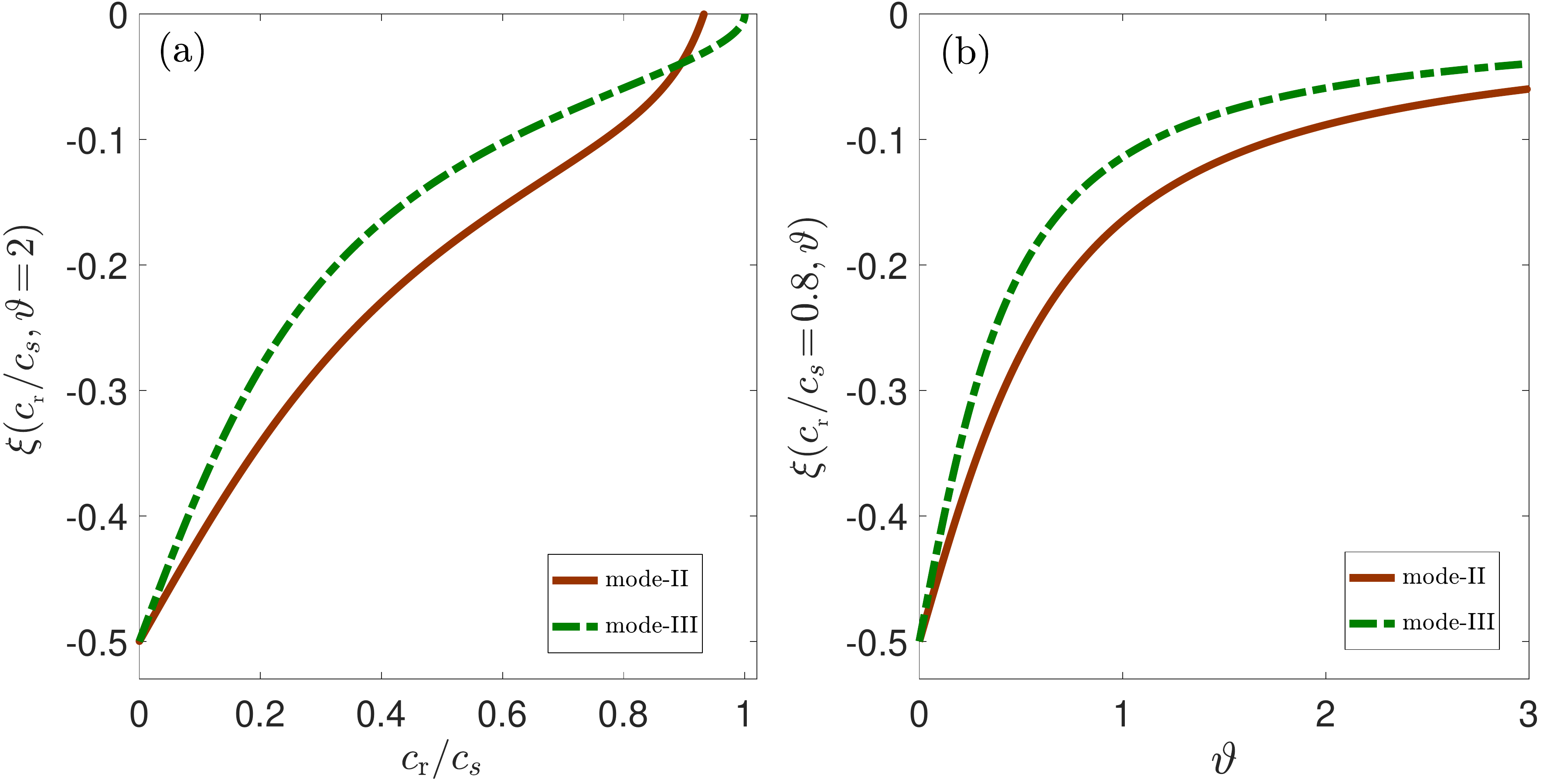}
\caption{(a) The unconventional singularity order $\xi$ vs. $c_{\rm r}/c_s$ with $\vartheta\!=\!2$ for mode-II (solid line), obtained from numerical solutions of Eq.~\eqref{eq:spectrum}, and for mode-III (dashed-dotted line), obtained from numerical solutions of Eq.~\eqref{eq:spectrum_modeIII}. In the mode-II calculations $c_d\!=\!2c_s$ was used, corresponding to $c_{_{\rm R}}\!\simeq\!0.9325c_s$. (b) The same as panel (a), but here $\xi$ is plotted against $\vartheta$, with $c_{\rm r}\!=\!0.8c_s$. See discussion of the presented results in the text.}
\label{fig:singularity power}
\end{figure}

In Fig.~\ref{fig:singularity power} we present $\xi(c_{\rm r}/c_s,\vartheta)$, obtained from numerical solutions of Eq.~\eqref{eq:spectrum}. In Fig.~\ref{fig:singularity power}a, we plot $\xi$ as function of $c_{\rm r}/c_s$ (solid line) for a fixed value of $\vartheta$ (using $c_d/c_s\=2$). The numerical solution exhibits all of the analytically predicted properties, i.e.~$\xi\!\to\!-\tfrac{1}{2}$ for $c_{\rm r}/c_s\!\to\!0$, a linear in $c_{\rm r}/c_s$ deviation from $\xi\=-\tfrac{1}{2}$ for small $c_{\rm r}/c_s$ and $\xi\!\to\!0$ for $c_{\rm r}\!\to\!c_{_{\rm R}}$ (in this case, for $c_d/c_s\=2$, we have $c_R\!\simeq\!0.9325c_s$). In Fig.~\ref{fig:singularity power}b, we plot $\xi$ as function of $\vartheta$ (solid line) for $c_{\rm r}/c_s\=0.8$ (again using $c_d/c_s\=2$). The numerical solution again agrees with the analytically predicted properties, i.e.~$\xi\!\to\!-\tfrac{1}{2}$ for $\vartheta\!\to\!0$, a linear in $\vartheta$ deviation from $\xi\=-\tfrac{1}{2}$ for small $\vartheta$ and an increase of $\xi$ toward $0$ with increasing $\vartheta$. All in all, these results demonstrate that unlike the conventional singularity case, the unconventional singularity order $\xi(c_{\rm r}/c_s,\vartheta)$ depends continuously on the interfacial constitutive relation, through $\eta$, and on the  crack propagation speed $c_{\rm r}$.

As long as only the singularity order $\xi$ is of interest, i.e.~if one is not interested in the off-interface fields of Eqs.~\eqref{eq:mode_II}, the interfacial integral representation discussed in relation to Eq.~\eqref{eq:boundary_integral} can be invoked. In particular, it yields
\begin{equation}
\label{eq:integral_again}
\eta\,v(r) = \frac{D(c_{\rm r})}{2\pi (c_{\rm r}/c_s)^3 \alpha_s(c_{\rm r})} \,\frac{\mu}{c_s}\int_0^\infty \frac{v(s)}{s-r}ds  \ ,
\end{equation}
which identifies with Eq.~\eqref{eq:boundary_integral} once the dimensionless function of $c_{\rm r}/c_s$ (the missing proportionality pre-factor therein) is read off and once the linear viscous-friction relation is substituted (neglecting the constant $\tau_{\rm res}$, which does not appear to leading order in the ``outer problem'' considered here). Invoking the ansatz $v(r)\!\sim\!r^\xi$, Eq.~\eqref{eq:integral_again} leads to the solution derived in Eq.~\eqref{eq:spectrum}, showing that the two approaches agree.

The results derived in Eqs.~\eqref{eq:mode_II},~\eqref{eq:spectrum} and~\eqref{eq:integral_again} apply for in-plane shear loading conditions (mode-II symmetry). The corresponding results for anti-plane shear loading conditions (mode-III symmetry) can be derived as well. Starting from the Navier-Lam\'e equation in Eq.~\eqref{Navier} and focusing on anti-plane shear deformation of the form ${\B u}(x,y,t)\=(0,0,u_z(x,y,t))$, one obtains  $\mu\nabla^2{u_z}\=\rho\,\ddot{u}_z$. Transforming to the crack edge co-moving polar coordinate system $(r,\theta)$ and demanding the continuity of the shear stress $\sigma_{zy}(x,y,t)\=\mu \partial_y u_z(x,y,t)$ across the frictional interface, one obtains the following solution using a standard complex-variable technique~\citep{Freund1998}
\begin{equation}
\label{eq:mode_III}
u_z(r,\theta)=\frac{2K^{(\xi)}_{\rm III}}{\mu\sqrt{2\pi}\,\alpha_s(c_{\rm r})} r_s^{\xi+1}\sin{\left[(\xi+1)\theta_s\right]}\ .
\end{equation}
This displacement solution is the anti-plane shear analog of Eqs.~\eqref{eq:mode_II}, where $K^{(\xi)}_{\rm III}$ generalizes the conventional mode-III stress intensity factor $K_{\rm III}$. Using then $v(r)\=2\dot{u}_z(r, \theta\=\pi)$, together with the viscous-friction boundary condition $\sigma_{zy}(r, \theta\=\pi)\=\eta\,v(r)$, one obtains
\begin{equation}
\cot(\xi\pi) = -\frac{2\,\vartheta\,(c_{\rm r}/c_s)}{\alpha_s(c{\rm_r})} \ .
\label{eq:spectrum_modeIII}
\end{equation}

Equation~\eqref{eq:spectrum_modeIII} is the anti-plane shear analog of Eq.~\eqref{eq:spectrum}. Its solution features the same qualitative properties of the solution of Eq.~\eqref{eq:spectrum} in terms of the dependence on $c_{\rm r}/c_s$ and $\vartheta$, though there are quantitative differences, as demonstrated in both panels of Fig.~\ref{fig:singularity power} (dashed lines). Notably, since the denominator in Eq.~\eqref{eq:spectrum_modeIII} vanishes when $c_{\rm r}\!\to\!c_s$, rather than when $c_{\rm r}\!\to\!c_{_{\rm R}}\!<\!c_s$ as in Eq.~\eqref{eq:spectrum}, $\xi(c_{\rm r}/c_s,\vartheta)$ in mode-III vanishes in the limit $c_{\rm r}\!\to\!c_s$ (cf.~Fig.~\ref{fig:singularity power}a). The results presented in Fig.~\ref{fig:singularity power} highlight the fact that unlike for the conventional singularity, which is independent of the symmetry mode of loading, the unconventional singularity order $\xi(c_{\rm r}/c_s,\vartheta)$ does depend on it (though the qualitative features are similar). Finally, the anti-plane shear analog of the interfacial integral representation in Eq.~\eqref{eq:integral_again} takes the form
\begin{equation}
\eta\,v(r) = \frac{\alpha_s(c_{\rm r})}{2\pi (c_{\rm r}/c_s)} \,\frac{\mu}{c_s} \int_0^\infty \frac{v(s)}{s-r}ds  \ .
\label{eq:integral_again_modeIII}
\end{equation}
The latter identifies with Eq.~\eqref{eq:boundary_integral} once the dimensionless function of $c_{\rm r}/c_s$ (the missing proportionality pre-factor therein) is read off. Invoking the ansatz $v(r)\!\sim\!r^\xi$, Eq.~\eqref{eq:integral_again_modeIII} leads to the solution derived in Eq.~\eqref{eq:spectrum_modeIII}, again showing that the two approaches agree. We note that one can generalize the interfacial integral eigenvalue problems defined in Eqs.~\eqref{eq:integral_again} and~\eqref{eq:integral_again_modeIII} to interfaces separating different materials~\citep{Weertman1980}, allowing in principle to derive the unconventional singularity order $\xi$ in the presence of material contrast, i.e.~for the so-called bi-material interfaces.

\section{Summary and concluding remarks}
\label{sec:summary}

In this paper we have shown that whenever the boundary conditions along crack surfaces are given as relations between different fields, and in particular for frictional shear cracks where the frictional strength generically depends on the slip velocity, unconventional singularities emerge. That is, near crack edge solutions quite generically feature unconventional singularities of order $\xi$, which is different from the conventional $-\tfrac{1}{2}$ singularity of classical cracks. A general theory of unconventional singularities has been developed, showing how $\xi$ can be calculated using an interfacial integral relation coupled to an interfacial constitutive law, corresponding to either the ``inner'' or ``outer'' problems.

The quite far-reaching implications of the existence of unconventional singularities for crack energy balance and dynamics have been discussed, giving rise to new physics compared to conventional fracture mechanics. It has been shown that $\xi\!\ne\!-\tfrac{1}{2}$ is accompanied by the breakdown of scale separation, which is at the heart of classical fracture mechanics. It implies that crack-associated dissipation is no longer restricted to a small region of size $\ell$ near the crack edge, but is rather spatially extended. It is quantified by the breakdown energy $E_{_{\rm BD}}\!(r)$, which differs from the edge-localized fracture energy $G_{\rm c}$ (quantifying the crack-associated dissipation on a scale $\ell$) and depends on $r\!>\!\ell$. Depending on the value of $\xi$ relative to $-\tfrac{1}{2}$, $E_{_{\rm BD}}\!(r)$ can be smaller or larger than $G_{\rm c}$.

In the presence of $\xi\!\ne\!-\tfrac{1}{2}$, the elastic energy influx into the edge region (the energy release rate) $G(r)$ also becomes scale dependent. Crack propagation energy balance can still be defined on a scale $r\!>\!\ell$ using $G(r)\=E_{_{\rm BD}}\!(r)$. The latter relation has been shown to be independent of $r$, but to explicitly depend on the edge dissipation length $\ell$, yet another qualitative difference compared to conventional fracture mechanics. The length $\ell$ can be meaningfully defined if the physics of crack edge dissipation is quite different from the physics of the spatially-extended dissipation. Such differences, if exist, manifest themselves both in the spatial distributions of the fields behind the propagating cracks and in the behavior of $G_{\rm f}(\delta)$, defined in Eq.~\eqref{eq:Gf}, which is commonly invoked in the context of earthquake physics. In the latter case, the existence of a well-defined edge-localized dissipation length $\ell$ implies that
$G_{\rm f}(\delta)$ changes its functional form quite significantly at a characteristic slip displacement $\delta_\ell$.

The general concepts and theory of unconventional singularities have been demonstrated for frictional shear cracks described by a linear viscous-friction constitutive law. Explicit calculations of $\xi\!\ne\!-\tfrac{1}{2}$, for both mode-II and mode-III symmetries, have been presented. $\xi$ has been shown to depend on the friction law (i.e.~on the crack surfaces boundary condition), on the propagation speed and on the symmetry mode of loading, all in qualitative difference to the classical theory of cracks, LEFM. Finally, we have presented asymptotic near edge unconventional singular solutions, for both mode-II and mode-III symmetries, going beyond the calculation of the unconventional singularity order $\xi$ alone.

We expect these results, beyond their general theoretical merit, to be useful for understanding frictional rupture in a wide variety of physical problems. These include quantifying and interpreting frictional dynamics in computer simulations, in laboratory experiments and in the context of geophysical observations. We hope that the theoretical tools developed in this paper will be indeed used in such contexts. In fact, they have been recently employed in interpreting measurements of breakdown energies in a wide variety of earthquakes~\citep{Viesca2015} and in developing a theory of frictional rupture that resolved various puzzling observations in extensive computer simulations~\citep{Brener2020unconventional}.

\vspace{0.5cm}

\noindent{\bf Acknowledgements}\\

We thank Robert Viesca for calling our attention to Ida's 1974 paper and his own 2015 paper with Garagash. E.B.~acknowledges support from the Israel Science Foundation (grants no.~295/16 and~1085/20), the Ben May Center for Chemical Theory and Computation, and the Harold Perlman Family.


\end{document}